# The criteria of order change in open system: the statistical approach

Viktor I. Shapovalov

The Volgograd branch of Moscow humanitarian-economical institute, Volgograd, Russia

shavi@rol.ru

The criteria determining the sign of entropy change in the open system are formulated. The concepts of entrostat, degree of openness, critical level of ordering are entered. The opportunity of occurrence of entropy oscillations in a stationary status is shown. The important role of considered of entropy laws in formation of the global tendencies.

**1. The criteria of entropy change in open system.**

The general remark: in the present manuscript as entropy $S$ is understood the statistical entropy of the following form $-k \int f(X) \ln f(X) dX$, $\int f(X) dX = 1$, $f(X)$ – function of distribution; all variables are the statistical ones.

1.1. Let $S[X]$ is an entropy of the closed system in an equilibrium state. Due to the external influence this system becomes open and the changes take place in it. The system comes to a stationary state at constant external influence. We shall designate as $S[X|Y]$ the entropy in a stationary state (here we use the designation of conditional entropy), $X$ – variable, describing a state of the system; $Y$ – variable, describing changes in the system which has arisen due to the external influence. The inequality is known

$$S[X] \geqslant S[X|Y] \quad . \tag{1}$$

The following concern to its basic lacks:

a) the inequality (1) does not allow to compare entropies of the different stationary states of open system;

b) the application of an inequality (1) to the problem of thermal contact of two objects forming isolated system, results in the contradiction (really, it follows from an inequality (1), that entropy of each of two objects after contact can not increase).

It will be shown below, that these lacks can be eliminated by introducing the entrostat concept.

1.2. We name as *the entrostat* the external environment not varying its own entropy at its influence at the system. In practice, the entrostat is any external environment satisfying to a condition

$$\frac{|\Delta S|}{S} \gg \frac{|\Delta S_e|}{S_e} \quad ,$$

where $\Delta S$ and $\Delta S_e$ – the entropy change of the researched system and the external environment caused by their interaction, accordingly.

The closest concept in physics to the entrostat is thermostat – the system with number of degrees of freedom aspiring to infinity.

1.3. At interaction of system with entrostat all changes occur only in the system. Hence, all variables of this interaction concern only to the system.

This rule is a basis for the further reasoning.

1.4. Let us compare two stationary states differing by the value of entrostat influence on the system. The first state differs from the closed one by changes, described by variables $Y_1, Y_2..., Y_{n-1}$, second state – by changes described by variables $Y_1, Y_2..., Y_n$. Let's write down expressions for appropriate conditional entropies:

$$S[X|Y_1...Y_{n-1}] = -k \iint ...\int f(XY_1...Y_{n-1}) \ln f(X|Y_1...Y_{n-1}) dX\, dY_1...dY_{n-1} \quad (2)$$

in the first state;

$$S[X|Y_1...Y_n] = -k \iint ...\int f(XY_1...Y_n) \ln f(X|Y_1...Y_n) dX\, dY_1...dY_n \quad (3)$$

in the second state.

Let's prove an inequality

$$S[X|Y_1...Y_{n-1}] > S[X|Y_1...Y_n] \quad . \quad (4)$$

Before proving we show, that it is possible to present (2) as

$$S[X|Y_1...Y_{n-1}] = -k \iint ...\int f(XY_1...Y_n) \ln f(X|Y_1...Y_{n-1}) dX\, dY_1...dY_n \quad . \quad (5)$$

Having taken a well-known expression of functions of distribution

$$f(XY_1...Y_n) = f(X|Y_1...Y_n) f(Y_1...Y_n) = f(Y_n|XY_1...Y_{n-1}) f(XY_1...Y_{n-1}) \quad , \quad (6)$$

we shall transform integral in (5) as follows

$$-k \iint ...\int f(Y_n|XY_1...Y_{n-1}) f(XY_1...Y_{n-1}) \ln f(X|Y_1...Y_{n-1}) dX\, dY_1...dY_n =$$

$$= -k \iint ...\int f(XY_1...Y_{n-1}) \ln f(X|Y_1...Y_{n-1}) dX\, dY_1...dY_{n-1} \int f(Y_n|XY_1...Y_{n-1}) dY_n =$$

$$= -k \iint ...\int f(XY_1...Y_{n-1}) \ln f(X|Y_1...Y_{n-1}) dX\, dY_1...dY_{n-1} \quad ,$$

that coincides with the right part of (2). Hence, (5) is correct.

With the account of (5) and (3) we write down and transform a difference:

$$S[X|Y_1...Y_{n-1}] - S[X|Y_1...Y_n] =$$

$$= k \iint ...\int f(XY_1...Y_n) \ln \frac{(f(X|Y_1...Y_n))}{(f(X|Y_1...Y_{n-1}))} dX\, dY_1...dY_n >$$

$$> k \iint ...\int f(XY_1...Y_n)[1 - \frac{(f(X|Y_1...Y_{n-1}))}{(f(X|Y_1...Y_n))}] dX\, dY_1...dY_n =$$

$$= k \iint ...\int f(XY_1...Y_n) dX\, dY_1...dY_n -$$

$$- k \iint ...\int f(Y_1...Y_n) f(X|Y_1...Y_{n-1}) dX\, dY_1...dY_n =$$

$$= k - k \iint \ldots \int f(Y_n|Y_1\ldots Y_{n-1}) f(Y_1\ldots Y_{n-1}) f(X|Y_1\ldots Y_{n-1}) dX\, dY_1\ldots dY_n =$$

$$= k - k \iint \ldots \int f(Y_n|Y_1\ldots Y_{n-1}) f(XY_1\ldots Y_{n-1}) dX\, dY_1\ldots dY_n =$$

$$= k - k \iint \ldots \int f(XY_1\ldots Y_{n-1}) dX\, dY_1\ldots dY_{n-1} \int f(Y_n|Y_1\ldots Y_{n-1}) dY_n =$$

$$= k - k = 0.$$

The well-known ratio was used in this transformation: $\ln a > 1 - 1/a$ ($a \neq 1$).

Thus, the inequality (4) is proved.

1.5. The inequality (4) can be written down in the expanded kind in view of the fact of different values of $n$:

$$S[X] > S[X|Y_1] > S[X|Y_1 Y_2] > \ldots > S[X|Y_1 Y_2 \ldots Y_i] > \ldots > 0 \quad . \tag{7}$$

This expression allows comparing entropies of stationary states of the open system.

It is important to mean, that (7) is fair only for systems interacting entrostat. The neglect by the specified circumstance results in the contradiction. In particular, it was marked in item 1.1. Really, in the problem of thermal contact of two objects it is impossible to consider any of them as entrostat. Therefore, in the given problem, the inequality (1), being a part of (7), can not be applied.

1.6. The transitions from one inequality to another in expression (7) occur due to the change of volume of entrostat influence at the system. The entrostat influence is considered to be constant during some interval of time, sufficient for system to reach a stationary state.

As *a degree of openness* $\alpha$ we shall name the phenomenological parameter quantitatively describing the value of entrostat influence on the system.

As it is visible from (7), each value of $\alpha$ unequivocally corresponds to the certain stationary value of entropy. The limiting states of the system occupy the extreme positions of a line (7). $\alpha = 0$ is carried out for an extreme left position (absolutely closed state), for extreme right one: $\alpha = \alpha_{max}$, (it is maximum opened state).

1.7. Let us introduce a designation: $S_{\alpha_i}$ – stationary value of entropy of the system having a degree of openness $\alpha_i$. Then (7) is possible to write as:

$$S_{\alpha=0} > S_{\alpha_1} > S_{\alpha_2} > \ldots > S_{\alpha_i} > \ldots > 0$$

The given expression evidently represents contained in (7) laws.

1). To reduce entropy in the system from $S_{\alpha_1}$ up to $S_{\alpha_2}$, it is necessary to increase an openness of system from $\alpha_1$ up to $\alpha_2$ (i.e. to increase entrostat influence). To increase entropy in system from $S_{\alpha_2}$ up to $S_{\alpha_1}$, it is necessary to reduce an openness of system from $\alpha_2$ up to $\alpha_1$ (i.e. to reduce entrostat influence).

2). The stationary entropy value $S_\alpha$ unequivocally corresponds to each degree of openness α. From here: a) if in system $S > S_\alpha$, the processes reducing entropy up to $S_\alpha$ should prevail in it; b) if $S < S_\alpha$, the processes increasing entropy up to $S_\alpha$ should prevail; c) if $S = S_\alpha$, the action of processes reducing and increasing entropy will compensate each other and the state of system becomes stationary.

3). If $\alpha = \alpha_{max}$, then $S_\alpha = 0$. Therefore in a non-stationary state $S > S_\alpha$ is carried out at each moment of time. Hence, in as much as possible opened system all processes should be accompanied by entropy reduction (see item "a" above).

1.8. Summarizing told above, we formulate *criteria of entropy change of system under influence of entrostat*:

a) at $\alpha = 0$ all processes are accompanied by increase of entropy of the system (well-known law of entropy increase);

b) at $\alpha = \alpha_{max}$ all processes are accompanied by entropy reduction (*law of entropy decrease*);

c) at $0 < \alpha < \alpha_{max}$, if $S > S_\alpha$, then the processes of entropy reduction prevail, if $S < S_\alpha$, then the processes of entropy increase prevail.

The information on some possible directions of use of the given criteria can be found in [1-3]. One of directions will be considered in the following section.

**2. Expansion of the Prigogine theorem on the minimal entropy production to the open systems under entrostat influence.**

2.1. Let us write down well-known equation of entropy balance as

$$\sigma = \frac{ds}{dt} = -\nabla \vec{J} + \sigma_i = \sigma_e + \sigma_i , \tag{8}$$

where σ – rate of the local entropy change; $s = dS/dV$ – local entropy; S – entropy of the system; V – volume; J – density of an entropy flow through surface area of perpendicular direction to the flow; $\sigma_e$ – part of the local entropy change of speed caused by interaction with external environment; $\sigma_i$ – production of the local entropy.

We need one more known value:

$$P = \int_V \sigma \, dV = \frac{dS}{dt} \tag{9}$$

– the rate of system entropy change.

According to (8), $P = P_e + P_i$, where $P_i = \int_V \sigma_i \, dV$ – the entropy production in the system.

Let's remind that the theorem of the minimal entropy production is formulated for linear processes and in mathematical terms is represented by the following inequality:

$$\partial P_i / \partial t \leqslant 0 \; , \tag{10}$$

where the sign "=" corresponds to the stationary state. The given inequality is a consequence that at $\sigma_i > 0$ the extremum of function $\sigma_i$ (and, hence, $P_i$) corresponds to the minimum.

The theorem of the minimal rate of the entropy change in linear processes instead of (10) for systems under entrostat influence is formulated below.

2.2. As it was already told in item 1.3 entropy of entrostat does not change during entrostat influence upon the system. With due regard for this fact for $\sigma$ (см. (8)) we write down a known in nonequilibrium thermodynamics expression:

$$\sigma = \sum_{j=1}^{n} X_j I_j \; , \quad X_j = \partial s / \partial a_j \; ; \quad I_j = da_j / dt$$

($a_j$ – parameter of the state).

Following [4], let us explore a system consisting of two vessels of different temperatures and composition of substance divided by a partition. The constant difference of temperatures is supported between vessels. Partition removing results in two flows in the system. The thermal flow $I_1$ corresponds to the generalized force $X_1$, arising because of a difference of temperatures. The diffusion flow $I_2$ corresponds to the generalized force $X_2$, arising due to unequal composition of substance. The force $X_1$ is constant in contrary of $X_2$, as it is caused by a constant difference of temperatures. The diffusion stops at the achievement of a stationary state by system and the flow $I_2$ becomes equal to zero. The environment supporting a constant difference of temperatures is in the given system as an entrostat.

Having taken the known equations for linear processes

$$I_j = \sum_k L_{jk} X_k \quad \text{и} \quad L_{jk} = L_{kj}$$

($L_{jk}$ – kinetics factors), we come to the formulations distinguished from similar ones in [4] only by $\sigma$ instead of $\sigma_i$. In particular,

$$\partial^2 \sigma / \partial X_2^2 = 2 L_{22} \; ; \quad (\partial \sigma / \partial X_2)_{st} = 0 \; .$$

The index "st" specifies that the value of term corresponds to the stationary state. As everyone can see $\sigma$ has an extremum in linear processes in a stationary state.

The entropy of the system under entrostat influence can both increase and decrease. It depends on the sign of change of a degree of an openness of the system (see items 1.7 and 1.8). Let's consider two cases.

1). Let us assume, that the degree of an openness of system has increased (for example, the difference of temperatures between vessels has been increased). According to the laws described in the item 1.7, the system will aspire to the state with smaller value of entropy: $\sigma < 0$ and $P < 0$. Let's make replacement of variables:
$$\tilde{\sigma} = -\sigma > 0 \ , \quad \tilde{P} = -P > 0 \ .$$
It is easy to see, that for new variables previous reasoning is in force. Really,
$$\tilde{\sigma} = -ds/dt = d(-s)/dt = \sum_j X_j I_j \ , \quad X_j = \partial(-s)/\partial a_j \ .$$
In result we come to the positive square-law form on $X_j$:
$$\tilde{\sigma} = L_{11} X_1^2 + 2 L_{21} X_1 X_2 + L_{22} X_2^2 > 0 \ .$$
Whence: $L_{22} > 0$. Hence, in a stationary state of the system extremum of functions $\tilde{\sigma}$ and $\tilde{P}$ is the minimum:
$$\partial \tilde{P}/\partial t \leq 0 \ . \tag{11}$$
The sign "=" corresponds to the stationary state.

Let's name $\tilde{P}$ as *a rate of reduction (decrease) of system entropy*. As it follows from (11) this rate for linear processes should decrease while nearing of system to a stationary state.

2). Let us assume, that the degree of an openness of system has decreased (for example, we have reduced a difference of temperatures between vessels). In this case system will aspire to the state with more value of entropy: $\sigma > 0$ and $P > 0$. The square-law form in expression for $\sigma$ is positive. It means that we come to the similar to (10) inequality by its form:
$$\partial P/\partial t \leq 0 \ .$$
We receive the following expression having united the given inequality with (11),
$$\partial |P|/\partial t \leq 0 \ . \tag{12}$$
The sign "=" corresponds to a stationary state.

2.3. The inequality (12) is mathematical expression of *the theorem on the minimal rate of entropy change of open system under entrostat influence*. Its formulation says: in linear processes (at constant value of entrostat influence) while nearing of system to a stationary state the rate of its entropy change decreases; thus a) if the system becomes more ordered, the rate of entropy decrease in it becomes less; b) if the system becomes less ordered, the rate of entropy increase in it becomes less.

In the following section we use the given theorem to prove an opportunity of occurrence of entropy oscillations around of a stationary state.

### 3. The entropy oscillations.

Let us add some function $F(P,S)$ to the left part of an inequality (12):

$$\partial P/\partial t + F(P,S) = 0 \qquad (13)$$

Let's notice, that the function $F$, added to $\partial P/\partial t$, should have the same sense, as $\partial P/\partial t$. As (13) is correct for linear processes, let us present $F$ as

$$F(P,S) = F(P_\alpha, S_\alpha) + \beta(P - P_\alpha) + \mu(S - S_\alpha) = \beta P + \mu S - \mu S_\alpha \;,$$

$$\beta = (\partial F/\partial P)_{P_\alpha} \;; \quad \mu = (\partial F/\partial S)_{S_\alpha} \;;$$

the index "α" specifies that value of term is undertaken in a stationary state, the degree of which openness is equal to α. Besides, it was taken into account that $F$ and $P$ are equal to zero in a stationary state.

In result (13) will accept a form of the equation of oscillations (see (9)):

$$\partial^2 S/\partial t^2 + \beta \partial S/\partial t + \mu S = \mu S_\alpha \;. \qquad (14)$$

It is easy to be convinced of the equation (14) a) at $\mu < 0$ has only unstable stationary decisions; b) at $0 < \mu \leq \beta^2/4$ has the stationary aperiodical decision; c) at $\mu > \beta^2/4$ has the stationary oscillation decision. The stationary decisions represent oscillations around $S_\alpha$. At $\beta > 0$ entropy oscillations are fading. At $\beta < 0$ amplitude of oscillation increases with current of time.

Thus, in linear processes there is a principal opportunity of occurrence of entropy oscillations around of a stationary state for systems under entrostat influence. It is possible to give a strict substantiation to this phenomenon only on the basis of an inequality (12). Really, the well-known inequality (10) does not permit to change a signum of $P_i$ and, hence, it is impossible to receive equation of oscillations on base of it.

According to the author's opinion the example of the entropy oscillations can be the following well-known phenomenon. A crystal periodically grows and melts in the process of crystallization due to the fluctuations of substance temperature. The amplitude of these processes fades by the end of crystallization.

In the following section we consider display of the entropy oscillations in other area of natural sciences.

### 4. Formation of the global tendencies.

(Written by materials published in [5, 6]).

4.1. Let us introduce some obvious designations:

$S_0$ – value of entropy in the beginning of some process;

$\Delta S_\alpha = S_\alpha - S_0$ – entropy change of the system having a degree of an openness α and which has reached a stationary state;

$\Delta S_{\alpha=0} = S_{\alpha=0} - S_0$ – entropy change of the absolutely closed system which has reached balance. For irreversible processes under the law of entropy increase

$$\Delta S_{\alpha=0} > 0 \ .$$

According to item 1.7, $S_{\alpha=0} > S_\alpha$. Let's make a difference

$$\Delta \tilde{S}_\alpha = S_\alpha - S_{\alpha=0} < 0 \ . \qquad (15)$$

It is easy to see, that

$$\Delta S_\alpha = \Delta S_{\alpha=0} + \Delta \tilde{S}_\alpha \ . \qquad (16)$$

Thus, in open system the entropy change $\Delta S_\alpha$ sums up from the positive $\Delta S_{\alpha=0}$ and negative $\Delta \tilde{S}_\alpha$. The expression (16) differs by this fact from well-known: $\Delta S = \Delta_i S + \Delta_e S$, where $\Delta_i S$ – the entropy produced inside the system, $\Delta_e S$ – outflow or inflow of entropy to the system from the outside [4]. Really, $\Delta_e S$ can be both negative and positive, while always $\Delta \tilde{S}_\alpha$ is less zero.

4.2. According to (16), all processes in system are divided into processes increasing and reducing entropy. It gives the basis to consider $\Delta \tilde{S}_\alpha$ as a quantitative measure of ordering (order) in system in a stationary state.

We shall name negative value $\Delta \tilde{S}_\alpha$ as *a critical level of ordering* in the system having a degree of an openness α. For absolutely closed system $\Delta \tilde{S}_\alpha = 0$; for as much as possible opened system $|\Delta \tilde{S}_\alpha|$ corresponds to a maximum.

As a result it is possible to change formulation of laws described in items 1.7 and 1.8 as follows:

1) If the system is ordered below critical level, then the processes increasing the order prevail in it; if the system is ordered upper critical level, then the processes reducing the order prevail in it (*this is the generalized law of entropy change*).

2) There is an unequivocal conformity between a degree of an openness of system and critical level of ordering ( $\alpha \Leftrightarrow \Delta \tilde{S}_\alpha$ ). In order to increase or to reduce $\Delta \tilde{S}_\alpha$ it is necessary to increase or to reduce a degree of openness α, accordingly.

4.3. Now let us explain the relation of told above to global problems, which mankind becomes to collide more and more often at last time.

By its statistical expression entropy is connected to the probability of events. In practice the action of entropy laws change the probability of events. In particular, the events promoting realizations of these laws occur more often others.

The Earth is a system open in relation to space. Therefore it has a critical level of ordering $\Delta \tilde{S}_\alpha$. If the Earth is ordered below $\Delta \tilde{S}_\alpha$, then the processes increasing the order in it ought to prevail; if above $\Delta \tilde{S}_\alpha$, then destructive processes should prevail. In the first case the mankind increases the order as a whole more than disorder while transforming the environmental world. Up to what time it can proceed? As long as creating, it will not exceed a $\Delta \tilde{S}_\alpha$ of the planet. In this case destructive processes will appear to prevail. Thus the surplus of the order created by mankind will be destroyed (or will be compensated by destructions in an environment). By the inertia it will be destroyed a little bit more, than it is necessary to be lowered up to a critical level. Below the critical level the processes of ordering will prevail, and the mankind will build again houses, partition off the rivers by dams, etc. After some time it will again exceed a critical level. The destroying processes will prevail again and eliminate the constructed objects, etc. In other words, the entropy oscillations described in the previous section will arise in the system "Earth".

The system being above critical level initiates a wide spectrum of processes capable to destroy order surplus in it. It is easy to understand that in the case of the Earth there are wars among these fastest processes. Let's notice, that for two last centuries the sole processes effectively braking total volume of human construction are the world wars. In other words, while the mankind is engaged in peace time in transformation of a nature, it inevitably comes nearer to a critical level, and consequently, to war and-or to such calamity of nature, which on scales and speed of destruction is comparable to war. It is possible to judge that on a planet the critical level of ordering is already exceeded by occurrence of the characteristic tendency: increasing of intensity of natural calamities, destroying climate change, aggravation of ecological crisis, appreciable increasing of probability of accidents, technogenical accidents, epidemics, social conflicts, local wars, and other events promoting the disorder. The well-known "hotbed effect", considered to be responsible for global warming, appears to be only a part of the specified tendency.

4.4. The understanding of the true reasons of the named tendency allows seeing ways of its prevention. Increase of an openness of a planet (for example, as a result of purposeful and wide development of space, of the Moon, Mars, etc.) would also raise value of its critical level that would result in prevalence of processes of ordering. And only then the ecological programs could effectively restore natural environment and the mankind would come to a condition of steady peace existence.

**Acknowledgments**

I thank N. V. Kazakov (The Volgograd State Technical University, Russia) for useful discussion of the present work.